\documentclass{ws-procs9x6}

\def\etal {{\it et al.}}

\def\beq{\begin{equation}}
\def\eeq{\end{equation}}

\def\na{\nabla}

\def\al{\alpha}
\def\be{\beta}

\def\ga{\gamma}

\def\ep{\epsilon}
\def\vp{\varepsilon}

\def\la{\lambda}

\def\si{\sigma}

\def\ta{\tau}

\def\Ga{\Gamma}

\begin{document}

\title{
RENORMALIZATION IN QED AND QFT WITH A\\
LORENTZ- AND CPT-VIOLATING BACKGROUND }

\author{ILYA L.\ SHAPIRO}

\address{Departamento de F\'{\i}sica, ICE,
Universidade Federal de Juiz de Fora\\
Juiz de Fora, Minas Gerais, 36036-900, Brazil\\
E-mail: shapiro@fisica.ufjf.br}

\begin{abstract}
The general features of renormalization and the renormalization group
in QED and in general quantum field theories
in curved spacetime with additional
Lorentz- and CPT-violating background fields are reviewed.
\end{abstract}

\bodymatter

\section{Introduction}

There is a growing interest in exploring
theories with weakly broken Lorentz and CPT symmetries. The
importance of such studies should be clear, especially for those
who believe that these important symmetries are exact. Indeed,
the unique way to confirm this belief is to assume that they
can be actually broken and verify the physical consequences
(see Ref.\ \refcite{CPTL-Review} for the latest results in this respect).
The general parametrization of all possible ways to break
Lorentz and CPT symmetries has been formulated by A. Kosteleck\'y
\etal\ in a series of papers (see, e.g., Ref.\ \refcite{CPTL-1}) and the
focus of attention is now naturally shifting to experimental
work with different manifestations of the theories with broken
symmetries.

An important aspect of theories with broken symmetries is
related to their renormalization structure. The reason is
that consistency at the quantum level can impose restrictions
on the classical theory and, eventually, restrict the space
of parameters for experimental verification. On the other hand,
such a study can be helpful in establishing the relation between
different parameters related to the breaking of Lorentz and CPT
symmetries. In what follows we discuss, in particular, the relation
between violations of symmetries in the matter and gravitational
sectors of the theory.

\section{Renormalization}

In the well-studied cases of external fields, such as
electromagnetic, metric (see, e.g., Ref.\ \refcite{book,PoImpo} for the review
and further references) and torsion \cite{torsi} one can establish the
form of possible counterterms by means of covariance and
other symmetries from one side and power counting on another
side. Naively one may think that these arguments will not work,
e.g., for QED with broken Lorentz and CPT symmetries, \cite{CPTL-1}
\begin{eqnarray}
S &=& \int d^4x\sqrt{-g}\,\Big\{\,
\frac{i}{2}\,\bar{\psi}\Ga^\mu D_\mu\psi
- \frac{i}{2}\, D^{\star}_\mu\bar{\psi}\Ga^\mu\psi
-\bar{\psi}\, M\,\psi -\frac{1}{4} F_{\mu\nu}\,F^{\mu\nu}
\nonumber\\
&&
\hskip 50pt
- \frac{1}{4}\, (k_F)_{\mu\nu\al\be}\,F^{\mu\nu}\,F^{\al\be}
+ \frac{1}{2}\,(k_{AF})^\rho \,\vp_{\rho\la\mu\nu}
\,A^\la\,F^{\mu\nu}\,\Big\} .
\label{1}
\end{eqnarray}
In the last formula $\Ga^\nu = \ga^\nu + \Ga_1^\nu$
and $M = m + M_1$, where
\beq
\Ga_1^\nu  =  c^{\mu\nu}\ga_\mu + d^{\mu\nu} \ga_5\ga_\mu
+ e^\nu + i\,f^\nu\ga_5
+\frac{1}{2}\,g^{\la\mu\nu}\si_{\la\mu} ,
\label{1.1}
\eeq
\beq
M_1  =  a_\mu\,\ga^\mu + b_\mu\,\ga_5\,\ga^\mu
+ i\, m_5\ga_5 + \frac{1}{2}\, H_{\mu\nu}\,\si^{\mu\nu} .
\label{1.2}
\eeq
All these parameters are experimentally proved
to be very weak \cite{CPTL-Review}.

The action (\ref{1})
is perfectly Lorentz and CPT invariant if we treat all new
external parameters as fields which transform according to
their Lorentz representations. The symmetries
get broken if we fix these fields, for instance, considering them
to be constant in a given reference frame. However, the
renormalization of the theory can be performed in one given
frame, and therefore one can safely treat all the fields
$a_\mu$, $b_\mu$, $m_5$, $H_{\mu\nu}$,
$c^{\mu\nu}$, $d^{\mu\nu}$, $e^\nu$, $f^\nu$, and $g^{\la\mu\nu}$
as being covariant. For the sake of generality we have
included in the action (\ref{1}) an external metric and,
in this way, may use the conventional approach based on
covariance.

Consider the renormalization of the theory (\ref{1}). It is
easy to note that the set of external fields includes
dimensionless ones in Eq.\ (\ref{1.1}) and dimensional ones in
Eq.\ (\ref{1.2}). Furthermore, the renormalization of terms
with dimensionless and dimensional terms are problems of
different level of complexity.

The form of counterterms constructed from the dimensional fields
(\ref{1.2}) are strongly restricted by the superficial degree
of divergence arguments. At the same time there are no
such arguments for the dimensionless fields (\ref{1.1}). Without
symmetry arguments it is not really possible to take care of
renormalization with these terms. Then, the theory with
dimensionless fields (\ref{1.1}) requires an infinite number
of counterterms, hence an infinite number of extra terms in
the classical action. However, the situation changes dramatically
if we remember that the Lorentz- and CPT-violating terms should
be very weak. This means we can safely restrict consideration
to the terms linear in Lorentz- and CPT-violating fields and
all necessary structures can be controlled by symmetry.

According to the arguments presented above, the action of
renormalizable
theory in curved space includes: (i) covariant
generalization of the standard expression, e.g., as given
in Eq.\ (\ref{1}) for QED and possible nonminimal terms;
(ii) vacuum terms depending on the metric and Lorentz- and
CPT-violating parameters, but not on matter fields.

The one-loop calculations have been performed for QED \cite{CPTL-2},
including in curved space \cite{CPTL} and they are in a perfect
agreement with the consideration presented above. For example, this
can be observed in the (incomplete) expression for the vacuum
divergences of the 1-loop effective action,
\begin{eqnarray}
\Ga^{(1)}_{\rm div} &=&
- \frac{\mu^\ep}{(4\pi)^2\,\ep}\,\int\,d^nx\,\sqrt{-g}\,
\Big\{ R_{\mu\al}\na_\rho\na_\ta k_F^{\ta\al\rho\mu}
- \frac{1}{6}R\na_\mu\na_\nu k_F^{\mu\nu}
\nonumber\\
&&
+ \frac{1}{3}R_{\mu\rho\al\be}\na^\be\na_\ta k_F^{\ta\mu\al\rho}
- \frac{1}{12}k_F^{\rho\la\mu\ta}RR_{\rho\la\mu\ta}
+ \frac12 k_F^{\al\rho\ta\la}R^\mu\mbox{}_{\rho\ta\la}R_{\mu\al}
\,\Big\},
\label{3}
\qquad
\end{eqnarray}
where we write $(k_F)^{\mu\la\nu}\mbox{}_\la = (k_F)^{\mu\nu}$.

The expression (\ref{3}) tells us something essential about
renormalization of the theories under discussion.
In the curvature-dependent low-energy vacuum sector of the theory
one could expect, for dimensional reasons, terms of the form
\cite{CPTL-grav}
\beq
S_{{\rm full}} = \int d^4x\sqrt{-g}\,
\left\{\phi R + \phi^{\mu\nu} R_{\mu\nu}
+ \phi^{\mu\nu\al\be} R_{\mu\nu\al\be}\right\}
+ S_{\rm HD},
\label{4}
\eeq
where the term $S_{\rm HD}$ represents possible higher derivative terms.
As we can see from Eq.\ (\ref{3}), the terms presented in Eq.\ (\ref{4}) are
indeed necessary if we consider a spacetime dependent parameter
$k_F^{\al\be\mu\nu}$. The introduction of the terms (\ref{4}) is
necessary already at the classical level, for otherwise quantum
corrections produce such terms with infinite coefficients which
cannot be removed by renormalization.

\section{Renormalization group}

An additional illustration of the relations between different
Lorentz- and CPT-violating parameters in the matter and gravity
sectors can be obtained within the renormalization group method.
The renormalization group equations of the new low-energy sector
parameters in the gravitational sector have the form \cite{CPTL}
\begin{eqnarray}
\mu\frac{d\phi}{d\mu} &=&
\frac{1}{96\pi^2}\,\na_\mu\na_\nu k_F^{\mu\nu},
\nonumber\\
\mu\frac{d\phi^{\mu\nu}}{d\mu} &=&
-\frac{1}{16\pi^2}\, \na_\al \na_\be k_F^{\al\mu\be\nu} ,
\nonumber\\
\mu\frac{d\phi^{\mu\nu\al\be}}{d\mu} &=&
-\frac{1}{48\pi^2}\,\na^\be\na_\la k_F^{\la\mu\al\nu}.
\qquad\qquad\qquad
\label{5}
\end{eqnarray}
These equations show that the order of magnitude of the
parameters $\phi$ cannot be too much smaller
than that of the variation of the parameter $k_F^{\al\mu\be\nu}$.
It may be interesting to perform a systematic analysis
of these kinds of relations. In fact, for some particular
cases the restrictions coming from renormalization group
equations can compete with the direct experimental bounds.

\section{Conclusions}

We presented a brief and mainly qualitative review of the
renormalization properties of quantum field theories
in curved spacetime in
the presence of Lorentz- and CPT-violating terms. The
study of quantum corrections is useful for formulating the
theory in a consistent way.

\section*{Acknowledgments}

This work was partially supported by CNPq, FAPEMIG and ICTP.

\end{document}